# A multiscale mathematical model of cancer, and its use in analyzing irradiation therapies

Benjamin Ribba*[1], Thierry Colin[2] and Santiago Schnell[3]

Address: [1]Institute for Theoretical Medicine and Clinical Pharmacology Department, Faculty of Medicine R.T.H Laennec, University of Lyon, Paradin St., P.O.B 8071, 69376 Lyon Cedex 08, France, [2]Mathématiques Appliquées de Bordeaux, CNRS UMR 5466 and INRIA futurs, University of Bordeaux 1, 351 cours de la liberation, 33405 Talence Cedex, France and [3]Indiana University School of Informatics and Biocomplexity Institute, 1900 East Tenth Street, Eigenmann Hall 906, Bloomington, IN 47406, USA

Email: Benjamin Ribba* - ribba@upcl.univ-lyon1.fr; Thierry Colin - colin@math.u-bordeaux.fr; Santiago Schnell - schnell@indiana.edu

* Corresponding author

## Abstract

**Background:** Radiotherapy outcomes are usually predicted using the Linear Quadratic model. However, this model does not integrate complex features of tumor growth, in particular cell cycle regulation.

**Methods:** In this paper, we propose a multiscale model of cancer growth based on the genetic and molecular features of the evolution of colorectal cancer. The model includes key genes, cellular kinetics, tissue dynamics, macroscopic tumor evolution and radiosensitivity dependence on the cell cycle phase. We investigate the role of gene-dependent cell cycle regulation in the response of tumors to therapeutic irradiation protocols.

**Results:** Simulation results emphasize the importance of tumor tissue features and the need to consider regulating factors such as hypoxia, as well as tumor geometry and tissue dynamics, in predicting and improving radiotherapeutic efficacy.

**Conclusion:** This model provides insight into the coupling of complex biological processes, which leads to a better understanding of oncogenesis. This will hopefully lead to improved irradiation therapy.

## Background

Mathematical models of cancer growth have been the subject of research activity for many years. The Gompertzian model [1,2], logistic and power functions have been extensively used to describe tumor growth dynamics (see for example [3] and [4]). These simple formalisms have been also used to investigate different therapeutic strategies such as antiangiogenic or radiation treatments [5].

The so-called linear-quadratic (LQ) model [6] is still extensively used, particularly in radiotherapy, to study damage to cells by ionizing radiation. Indeed, extensions of the LQ model such as the 'Tumor Control Probability' model [7] are aimed at predicting the clinical efficacy of radiotherapeutic protocols. Typically, these models assume that tumor sensitivity and repopulation are constant during radiotherapy. However, experimental evidence suggests that cell cycle regulation is perhaps the most important determinant of sensitivity to ionizing radiation [8]. It has been suggested that anti-growth signals such as hypoxia or the contact effect, which are responsible for decreasing the growth fraction, may play a crucial role in the response of tumors to irradiation [9].

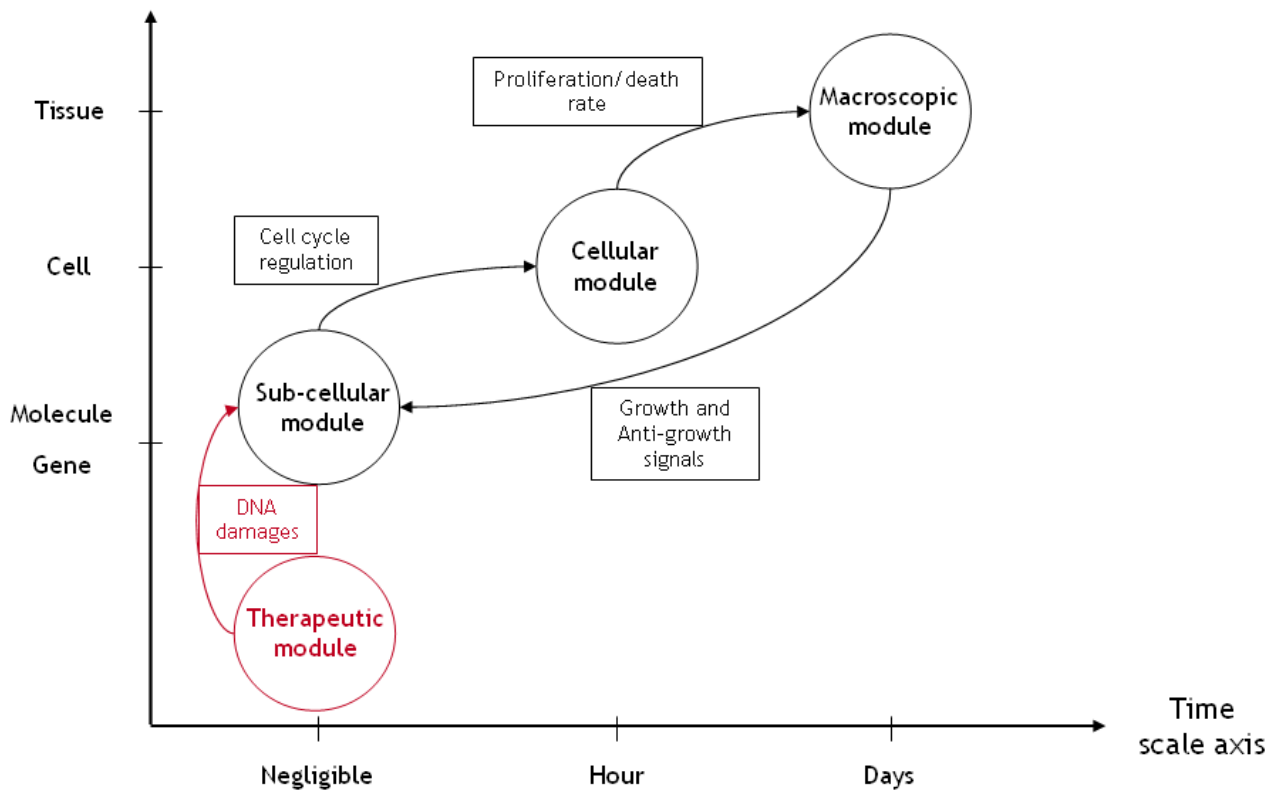

**Figure 1**
**Multiscale nature of the model**. Schematic view of the multiscale nature of the model, composed of four different levels. At the genetic level we integrate the main genes involved in the evolution of colorectal cancer within a Boolean network and this results in cell cycle regulation signals. The response to these signals occurs at the cellular level, determining whether each cell proliferates or dies. Given this information, the macroscopic model the new spatial distribution of the cells is computed at the tissue level. The number and spatial configuration of cells determine the activation of the antigrowth signals, which in turn is input to the genetic level. Irradiation induces DNA breaks, which, in the model, activate the *p53* gene at the genetic level.

Nowadays, computational power allows us to build mathematical models that can integrate different aspects of the disease and can be used to investigate the role of complex tumor growth features in the response to therapeutic protocols [10]. In the present study we propose a multiscale model of tumor evolution to investigate growth regulation in response to radiotherapy. In our model, key genes in colorectal cancer have been integrated within a Boolean genetic network. Outputs of this genetic model have been linked to a discrete model of the cell cycle where cell radiosensitivity has been assumed to be cycle phase specific. Finally, Darcy's law has been used to simulate macroscopic tumor growth.

The multiscale model takes into account two key regulation signals influencing tumor growth. One is hypoxia, which appears when cells lack oxygen. The other is overpopulation, which is activated when cells do not have sufficient space to proliferate. These signals have been correlated to specific pathways of the genetic model and integrated up to the macroscopic scale.

## Methods

Oncogenesis is a set of sequential steps in which an interplay of genetic, biochemical and cellular mechanisms (including gene pathways, intracellular signaling pathways, cell cycle regulation and cell-cell interactions) and environmental factors cause normal cells in a tissue to develop into a tumor. The development of strategies for treating oncogenesis relies on the understanding of patho-genesis at the cellular and molecular levels. We have therefore developed a multiscale mathematical model of these processes to study the efficacy of radiotherapy. Several mathematical frameworks have been developed to model avascular and vascular tumor growth (see [11-14]). Here we propose a multiscale mathematical model for avascular tumor growth, which is schematically presented in Figure 1. This model provides a powerful tool for addressing questions of how cells interact with each other and their environment. We use the model to study tumor regression during radiotherapy.

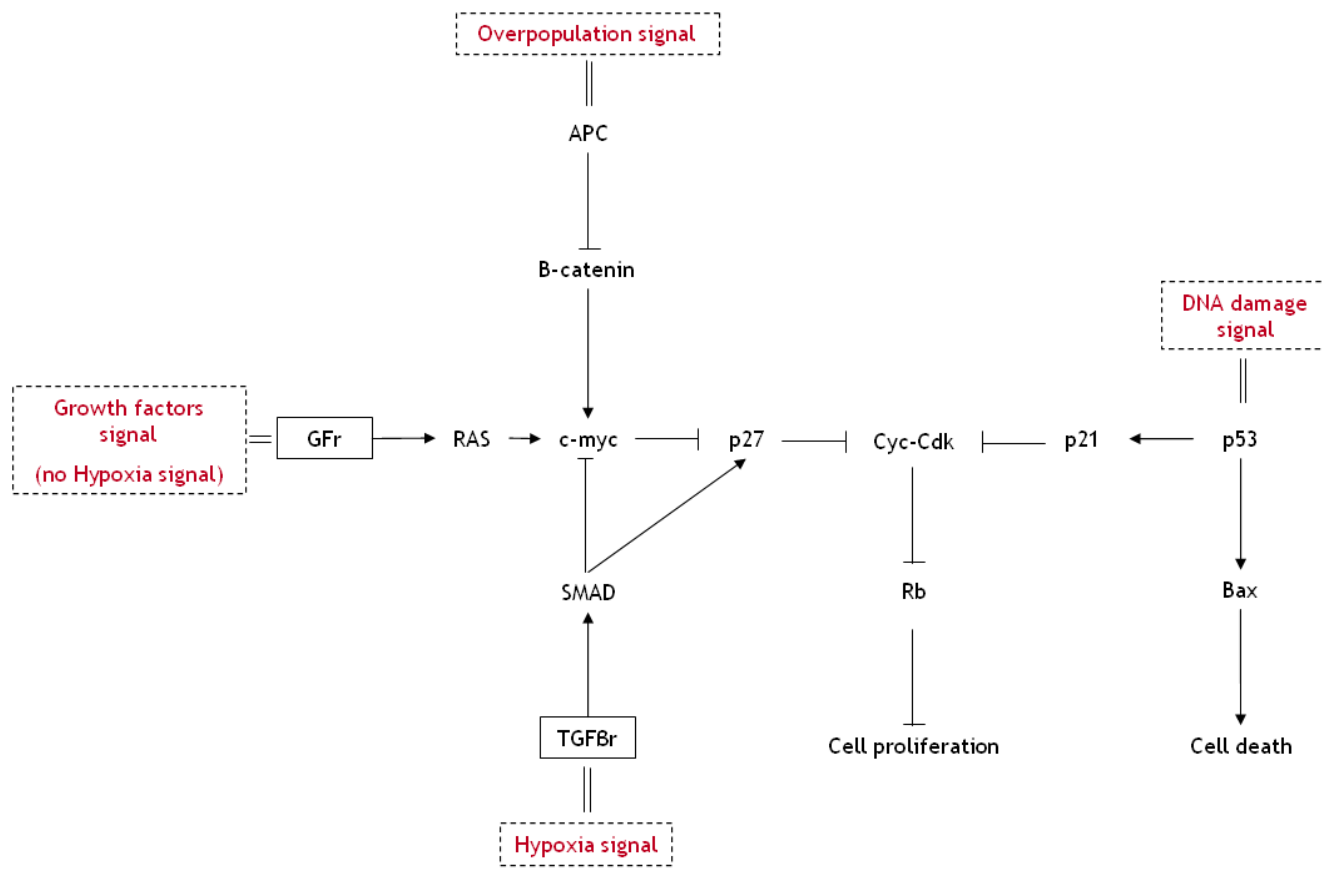

**Figure 2**
**Cell proliferation and death (genetic regulation) for colorectal cancer**. This figure shows the genetic model with regulation signals as inputs. *p53* is activated when DNA is damaged and leads the cell to apoptosis. *SMAD* is activated through *TGFβ* receptors during hypoxia and inhibits cell proliferation. Overpopulation inhibits cell proliferation through activation of *APC*. *RAS* promotes cell proliferation through growth factor receptors when sufficient oxygen is available for the cell, that is, there is no hypoxia. This flow chart was developed from knowledge available from bibliographic resources [15,16] and from the Knowledge Encyclopedia of Genes and Genomes [53,54].

*Gene level*
Five genes are commonly mutated in colorectal cancer patients, namely: *APC* (Adenomatosis Polyposis Coli), *K-RAS* (Kirsten Rat Sarcoma viral), *TGF* (Transforming Growth Factor), *SMAD* (Mothers Against Decapentaplegic) and *p53* or *TP53* (Tumor Protein 53). These genes belong to four specific pathways, which funnel external or internal signals that cause cell proliferation or cell death (see [15] and [16,17] for more details).

The anti-growth, *p53*, pathway is activated in the case of DNA damage [18,19]. This is particularly relevant during irradiation [20]. *p53* pathway activation can block the cell cycle and induce apoptosis [21,22]. The *K-RAS* gene belongs to a mitogenic pathway that promotes cell proliferation in the presence of growth factors [23]. Activation of the anti-growth pathways *TGFβ/SMAD* and *WNT/APC* inhibits cell proliferation. The *SMAD* gene is activated by hypoxia signals [24,25], while *APC* is activated through *β*-catenin by loss of cell-cell contact [26-30]. Moreover, it has recently been hypothesized that overpopulation of *APC* mutated cells can explain the shifts of normal proliferation in early colon tumorigenesis [31].

**Table 1: Apoptotic activity.** Apoptotic activity induced by two 20 *Gy* radiotherapy protocols applied to *APC*-mutated tumor cells.

| | Apoptotic activity | | | |
| --- | --- | --- | --- | --- |
| | Total dose (*Gy*) | Scheduling | Apoptotic fraction – mean – (%) | Apoptotic fraction – max – (%) |
| Standard protocol | 20 | 2 *Gy* daily | 2.59 | 4 |
| Heuristic | 20 | 2 *Gy* Repeated 10 times before hypoxia | 3.14 | 4.25 |

We assume that activation of *APC* and *SMAD* is due to overpopulation and hypoxia signals respectively. Both pathways inhibit cell proliferation. In consequence, *APC* mutated cells promote overpopulation and *SMAD* or *RAS* mutated cells promote proliferation during hypoxia. Figure 2 shows the schematic genetic model.

We develop a Boolean model of these pathways in Figure 2. Each gene is represented by a node in the network and the interactions are encoded as the edges. The state of each node is 1 or 0, corresponding to the presence or absence of the genetic species. The state of a node can change with time according to a logical function of its state and the states of other nodes with edges incident on it [32-34]. The rules governing the genetic pathways are presented in Table 2.

### Cell level

We consider a discrete mathematical model of the cell cycle in which the cycle phase duration values were set according to the literature [35]. In our model the proliferative cycle is composed of three distinct phases: *S* (DNA synthesis), $G_1$ (Gap 1) and $G_2M$ (Mitosis). We model the 'Restriction point' *R* [36] at the end of $G_1$ where internal and external signals, i.e. cell DNA damage, overpopulation and hypoxia, are checked [37] (see Figure 3 for a schematic representation of our cell cycle model).

For each spatial position *(x, y)*, we assume that:

- If the local concentration of oxygen is below a constant threshold $Th_o$ and if *SMAD* is not mutated, hypoxia is declared and causes cells to quiesce ($G_0$) through *SMAD* gene activation (see Figure 2);

- If the local number of cells is above a constant threshold $Th_t$ and if *APC* is not mutated, overpopulation is declared and leads cells to quiesce ($G_0$) through the *APC* gene (see Figure 2);

- Otherwise, if the conditions are appropriate, cells enter $G_2M$ and divide, generating new cells at the same spatial position.

Induction of apoptosis through *p53* gene activation is discussed later.

### Tissue level

We use a fluid dynamics model to describe tissue behavior. This macroscopic-level continuous model is based on Darcy's law, which is a good model of the flow of tumor cells in the extracellular matrix [38-40]:

$$v = -k\nabla p \quad (1)$$

**Table 2: Genetic model.** Boolean (logical) functions used in the genetic model depicted Figure 1. For *APC*, *SMAD* and *RAS*, Boolean values are set to 0, 0 and 1 respectively when genes are mutated.

| Boolean model | |
| --- | --- |
| Node | Boolean updating function |
| $APC^t$ | $APC^{t+1} = \begin{cases} 1 & \text{if Overpopulation signal} \\ 0 & \text{otherwise} \end{cases}$ |
| | $APC^{t+1} = 0$ *if mutated* |
| $\beta cat^t$ | $\beta cat^{t+1} = \neg APC^t$ |
| $cmyc^t$ | $cmyc^{t+1} = RAS^t \land \beta cat^t \land \neg SMAD^t$ |
| $p27^t$ | $p27^{t+1} = SMAD^t \lor \neg cmyc^t$ |
| $p21^t$ | $p21^{t+1} = p53^t$ |
| $Bax^t$ | $Bax^{t+1} = p53^t$ |
| $SMAD^t$ | $SMAD^{t+1} = \begin{cases} 1 & \text{if Hypoxia signal} \\ 0 & \text{otherwise} \end{cases}$ |
| | $SMAD^{t+1} = 0$ *if mutated* |
| $RAS^t$ | $RAS^{t+1} = \begin{cases} 1 & \text{if no Hypoxia signal} \\ 0 & \text{otherwise} \end{cases}$ |
| | $RAS^{t+1} = 1$ *if mutated* |
| $p53^t$ | $p53^{t+1} = \begin{cases} 1 & \text{if DNA damage signal} \\ 0 & \text{otherwise} \end{cases}$ |
| | $p53^{t+1} = 0$ *if mutated* |
| $CycCDK^t$ | $CycCDK^{t+1} = \neg p21^t \land \neg p27^t$ |
| $Rb^t$ | $Rb^{t+1} = \neg CycCDK^t$ |

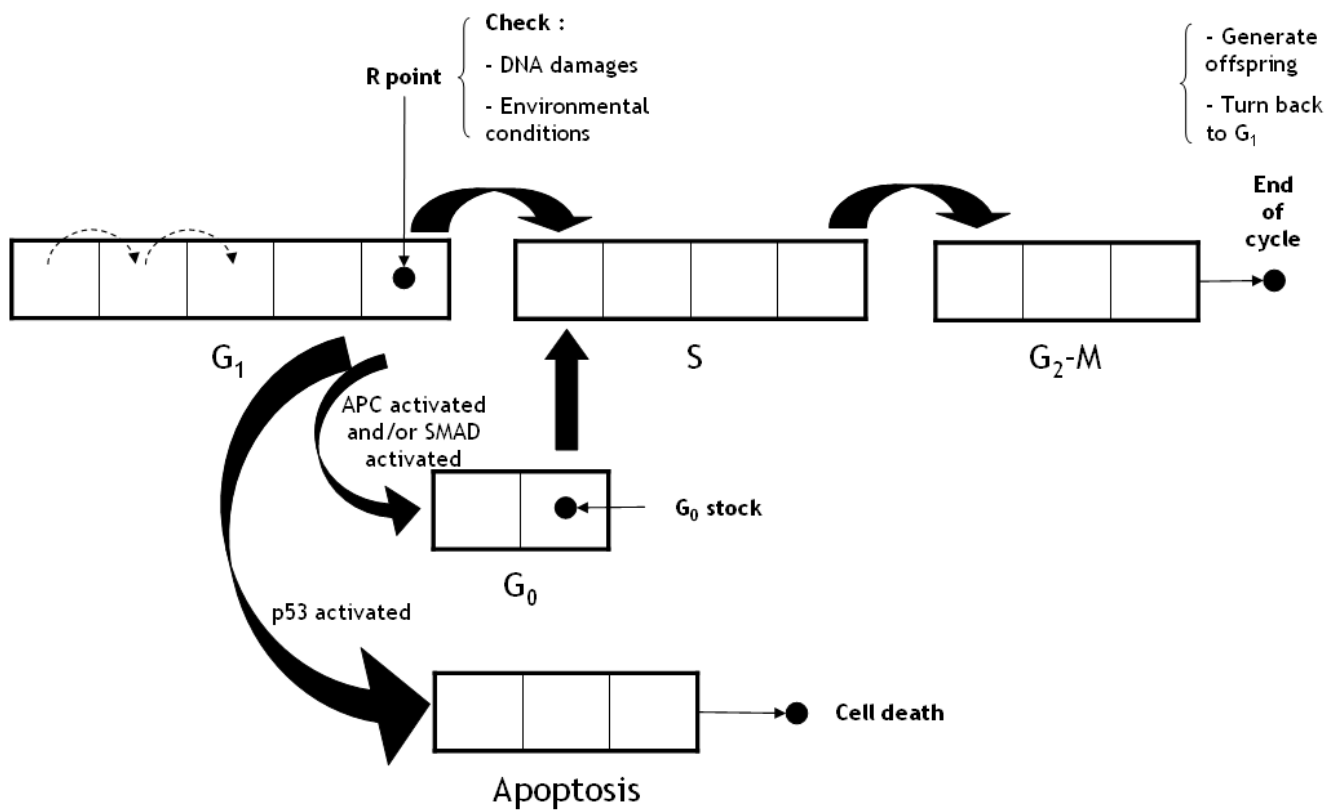

**Figure 3**
**Diagram of the cell cycle model**. In this discrete model, cells progress through a cell cycle comprising three phases: $G_1$, $S$, and $G_2M$. At the end of the $G_2M$ phase, cells divide and new cells begin their cycle in $G_1$. At the last stage of phase $G_1$, we modelled the restriction point $R$, where DNA integrity and external conditions (overpopulation and hypoxia) are checked. If overpopulation occurs, *APC* is activated; if hypoxia occurs, *SMAD* is activated. Both these conditions lead cells to $G_0$ (quiescence). Cells remain in the quiescent phase in the absence of external changes, otherwise they may return to the proliferative cycle (at the first step of $S$ phase). DNA damage can also activate the *p53* pathway, which leads cells to the apoptotic phase. Cells at the end of the apoptotic phase die and disappear from the computational domain.

where $p$ is the pressure field. The media permeability $k$ is assumed to be constant.

We study the evolution of the cell densities in two dimensions. We formulate the cell densities in the tissue mathematically as advection equations, where $n_\phi(x, y, t)$ represents the density of cells with position $(x, y)$ at time $t$ in a given cycle phase $\phi$. Assuming that all cells move with the same velocity given by Eq. (1) and applying the principle of mass balance, the advection equations are:

$$\frac{\partial n_\varphi}{\partial t} + \nabla \cdot (v n_\varphi) = P_\varphi \quad \forall \varphi \in \{G_1, S, G_2M, G_0, Apop\} \quad (2)$$

where $P_\phi$ is the cell density proliferation term in phase $\phi$ at time $t$, retrieved from the cell cycle model.

The global model is an age-structured model (see Section 2.7). Initial conditions for $n_\phi$ are presented in Section 2.6.

Assuming $\sum_\varphi n_\varphi$ to be a constant and adding Eq. (2) for all phases, the pressure field $p$ satisfies:

$$-\nabla \cdot (k \nabla p) = \sum_\varphi P_\varphi. \quad (3)$$

The pressure is constant on the boundary of the computational domain.

In our model, the oxygen concentration $C$ follows a diffusion equation with Dirichlet conditions on the edge of the computation domain $\Omega$:

$$\frac{\partial C}{\partial t} - \nabla \cdot (D \nabla C) = -\sum_\varphi \alpha_\varphi n_\varphi \quad on \quad \Omega/\Omega_{bv} \quad (4)$$

$$C = C_{\max} \quad on \quad \Omega_{bv} \quad (5)$$

**Table 3: Table of parameters** Table of numerical parameters used for simulations.

| | Model parameters | | | |
|---|---|---|---|---|
| Parameter | Description | Unit | Value | Reference |
| $T_{G_1}$ | Duration of $G_1$ phase | h | 20 | [35,44] |
| $T_S$ | Duration of S phase | h | 10 | [35,44] |
| $T_{G_2M}$ | Duration of $G_2$M phase | h | 3 | [35,44] |
| $T_{G_0}$ | Duration of $G_0$ phase | h | 5 | Estimated |
| $T_{Apoptosis}$ | Duration of the apoptotic phase | h | 5 | Estimated |
| $C_{max}$ | Oxygen in blood | $mlO_2$ | $10^{-2}$ | Estimated |
| $\alpha_\phi$ | Oxygen consumption in phase $\phi$ | $mlO_2 s^{-1}$ | $5 - 10 \times 10^{-15}$ | Estimated |
| $Th_o$ | Hypoxia threshold | $cell^{-1}$ | $5 \times 10^{-15}$ | Estimated |
| $Th_t$ | Overpopulation threshold | cell | 2000 | Estimated |
| $R_\phi$ | Cell Radio-sensitivity in phase $\phi$ | $Gy^{-1}$ | $0.2 - 2$ | [41-43] |
| k | Media permeability | $m^2$ | 0.2 | Estimated |

$C_{\partial\Omega} = 0$ (6)

*D* is the oxygen diffusion coefficient, which is constant throughout the computation domain. In this equation, $\Omega_{bv}$ stands for the spatial location of blood vessels, $\alpha_\phi$ is the coefficient of oxygen uptake by cells at cell cycle phase $\phi$ and $C_{max}$ is the constant oxygen concentration in blood vessels.

### Therapy assumptions

Cell sensitivity depends on cell cycle phase [8]. We assume that only proliferative cells are sensitive to the treatment. In addition, we assume that DNA damage is proportional to the irradiation dose. This is known as the 'single hit' theory, which is governed by the expression

$n_{dsb} = R_\phi d$ (7)

where $n_{dsb}$ is the number of double strand breaks induced by radiation dose *d*. As mentioned previously, the radiosensitivity $R_\phi$ has been assumed to depend on the cell cycle phase (see Table 3). Based upon radiobiological experiments found in the literature, we take the radiosensitivity as constant (2 $Gy^{-1}$) in $G_1$ and $G_0$. It decreases in S phase to 0.2 $Gy^{-1}$, and then increases to 2 $Gy^{-1}$ during $G_2$.

We set a constant treatment threshold $Th_r$ such that if $n_{dsb}$ due to the irradiation dose is above $Th_r$ at any time, p53 is activated and the cells are labeled as 'DNA damaged cells'. DNA damaged cells are identified at the R point of the cell cycle and are directed to apoptosis. They die and disappear from the computational domain after $T_{Apoptosis}$, i.e. the duration of the apoptotic phase.

The standard radiotherapy protocol used in the simulations consists of a 2 *Gy* dose delivered each day, five days a week, and can be repeated for several weeks. The radiotherapeutic dose is assumed to be uniformly distributed over the spatial domain.

According to the radiosensitivity parameters found in the literature [41-43], only a fraction of mitotic cells are assumed to be sensitive to the standard 2 *Gy* dose.

### Model parameters

Cell cycle kinetic parameters were retrieved from flow cytometric analysis of human colon cancer cells [35,44]. Table 3 summaries the quantitative parameters used in our model.

### Computational domain and initial conditions

In our two-dimensional model we study an 8 cm square tissue. We assume that the domain comprises five small circular tumor masses, the first located at the center of the computational domain and the other four towards the corners. Moreover, the domain has two sources of oxygen, to the right and left sides of the central cell cluster (see Figure 4).

The number of cells in each tumor is the same, and they are uniformly distributed. The number of cells in each phase of the cell cycle is proportional to the duration of the phase. For instance, the $G_1$ phase contains twice as many cells as the S phase because the $G_1$ phase is twice as long as the S phase. It is important to emphasize that the cell cycle phases are discrete (see Section 2.7).

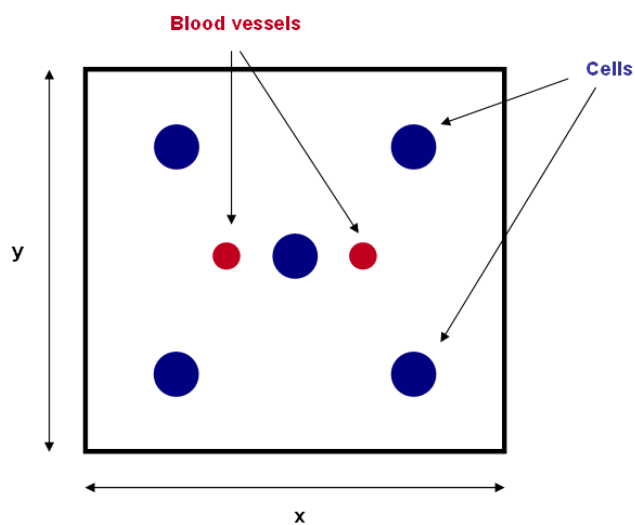

**Figure 4**
**Initial conditions**. Schematic representation of the two-dimensional computation domain for model simulations, with the initial spatial configuration of the cells. The domain is composed of five cell clusters and two blood vessels.

*Simulation technique*
The model is fully deterministic. Cell cycle phases durations $\tau_\phi$ have been discretized in several elementary age intervals $a \in \{1, ..., N_\phi\}$ where $N_\phi$ is an integer such as $\tau_\phi = dt \times N_\phi$. Here $dt$ is the time step of the cell cycle model. The cell density $n_{a,\phi}$ at age $a$ in phase $\phi$ is governed by:

$$\frac{\partial n_{a,\varphi}}{\partial t} + \nabla \cdot (v n_{a,\varphi}) = P_{a,\varphi}. \quad (8)$$

In this equation, $\phi \in \{G_1, S, G_2M, G_0, Apoptosis\}$ and $a \in \{1, ..., N_\phi\}$. $P_{a,\phi}$ is the cell density proliferation term in phase $\phi$ at age $a$ retrieved from the cell cycle model. In the simulations, the intracellular and extracellular conditions were identified for cells at the end of $G_1$ phase. These were used as initial conditions for the gene level model. The genetic model was computed until it reached steady state (this is of the order of 10 iterations).

Noting that $\sum_{a,\varphi} n_{a,\varphi}$ is constant, we can sum Eqs. (8) to obtain an expression for the pressure field of the form:

$$-\nabla \cdot (k \nabla p) = \sum_{a,\varphi} P_{a,\varphi}. \quad (9)$$

The computer program starts from an initial distribution of cells in each state $\{a, \phi\}$. The computations are performed using a splitting technique. First we run the cell cycle model for one time-step $dt$, then retrieve new values for $n_{a,\phi}$ and compute $P_{a,\phi}$. Pressure is retrieved by solving Eq. (9) and velocity is computed using Darcy's law (see Eq. (1)). Since the contribution of the source term has been taken into account by the cell cycle model at the first stage of the splitting technique, Eqs. (8) are solved continuously and without second members:

$$\frac{\partial n_{a,\varphi}}{\partial t} + \nabla \cdot (v n_{a,\varphi}) = 0, \quad (10)$$

which can also be written [using (9)]:

$$\frac{\partial n_{a,\varphi}}{\partial t} + v \cdot \nabla n_{a,\varphi} = \left( \sum_{a',\varphi'} P_{a',\varphi'} \right) n_{a,\varphi}. \quad (11)$$

This equation is then solved using a splitting technique. The advection parts of Eq. (11) are solved by sub-cycling finite different scheme computations, with time-step $dt_{adv}$ being smaller than $dt$ (for stability reasons). We set $n_{a,\phi} = 0$ on the part of the boundary where $v \cdot \upsilon < 0$, $\upsilon$ denoting the outgoing normal to the boundary. For the pressure $p$, we set $p = 0$ on the boundary.

All simulations (except the ones shown in Figure 7) were run for 320 h with time step $dt = 1$ h in a discrete computational domain composed by 100 × 100 elementary spatial units.

**Results and discussion**
We divide our results and discussion into three parts. The first section concerns simulations of the model without therapeutic interactions (Sections 3.1–3.2). The second part deals with the interactions between tumor growth and the effect of therapeutic protocols (Section 3.3). Finally, we investigate the sensitivity of the results to model parameters and initial conditions (Section 3.4). Genetic mutations are simulated by running the model, having set the Boolean values of particular genes constant (see Table 2). Since the genetic model is run until steady state is reached, simulation of mutated cell growth is equivalent to simulation of cells that are not sensitive to particular anti-growth signals. In the following, we will refer to cells with at least one mutation as 'cancer cells'. Cells with no mutations are called 'normal cells'.

*Gene-dependent tumor growth regulation*
Figure 5 shows the simulated growth of cell colonies. According to the model settings, the colony of normal cells grows up to $10^6$ cells and is then regulated through activation of gene *APC* owing to overpopulation. *APC* mutated tumor cells are not sensitive to overpopulation and reproduce exponentially until late regulation because of hypoxia, through *SMAD* gene activation. Finally, according to the model parameters, *APC* and *SMAD/RAS*

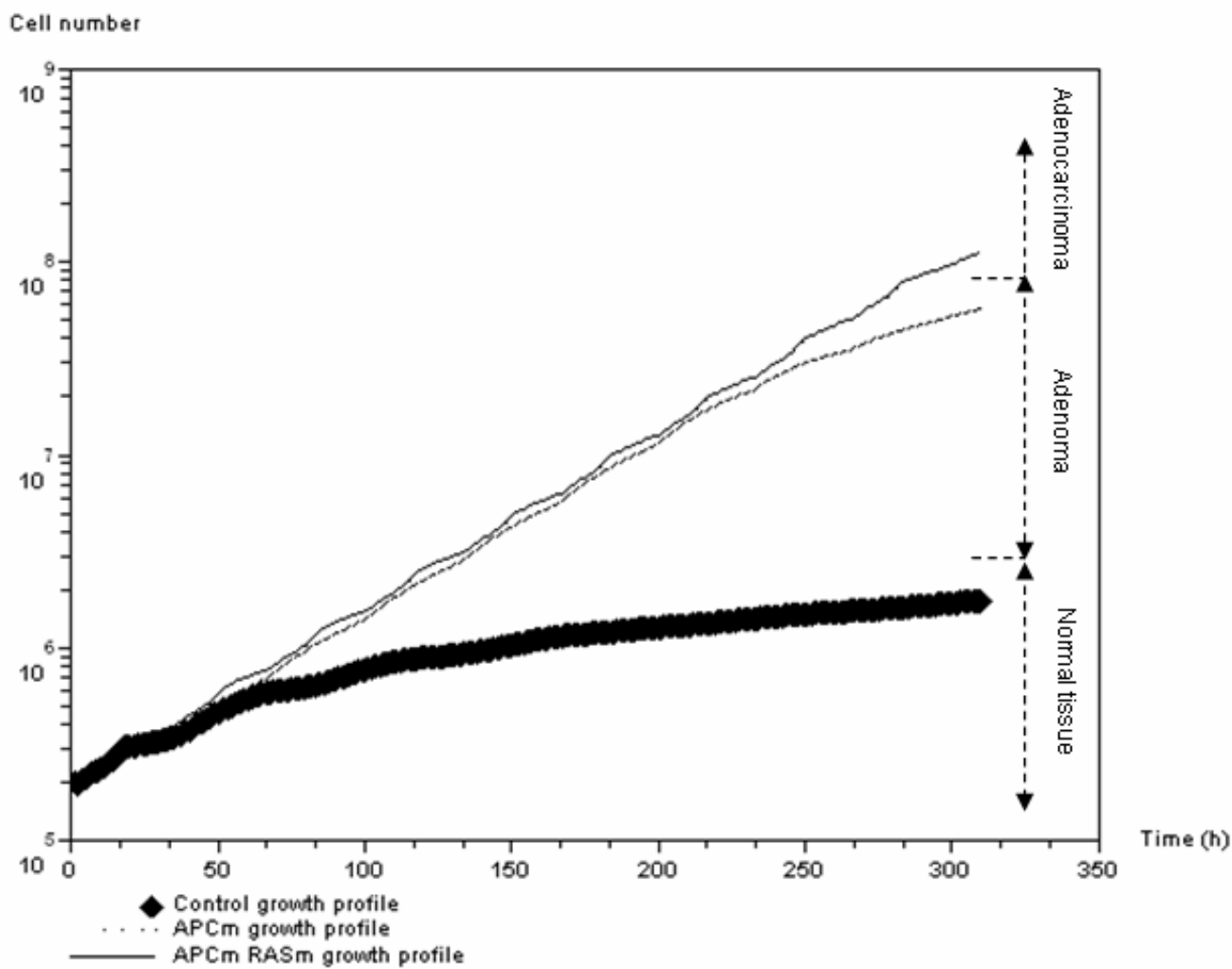

**Figure 5**
**Cell population growth**. Cell population growth (log plot) over time according to three different genetic profiles: normal cells (black diamonds), *APC* mutated cells (dashed line), and *APC* + *SMAD/RAS* mutated cells.

mutated tumor cells cannot be regulated at all and thus induce an exponential growth profile.

The simulation results reproduce the evolution of colorectal cancer [16,45]. Indeed, APC has been shown to promote shifts in pattern of the normal cell population in early colorectal tumorigenesis, and *SMAD/RAS* mutations promote evolution from early adenoma to adenocarcinoma.

### Features of anti-growth signals and effect on tumor growth
*APC-dependent growth regulation*
The top diagram of Figure 6 shows the evolution of the total and quiescent cell numbers, when population growth is regulated through activation of the *APC* gene due to overpopulation. Figure 6 shows that the first 100 hours are characterized by oscillations in both populations, which slowly disappear and become linear growth. Indeed, as the cell population begins to grow, it tends to activate *APC* signaling owing to overpopulation in the inner part of the tumor masses. This results in a rapid increase in the number of quiescent cells, which in turn slows cell proliferation. Cell advection leads to invasion of new tissues, which promotes proliferation and in turn slows the evolution of the quiescent cell population. These oscillations in cell population are caused by a combination of overpopulation signal propagation in the inner parts of the cell clusters and the cells' ability to move to colonize free space. Very soon, what was once free space becomes overpopulated. This results in a constant propor-

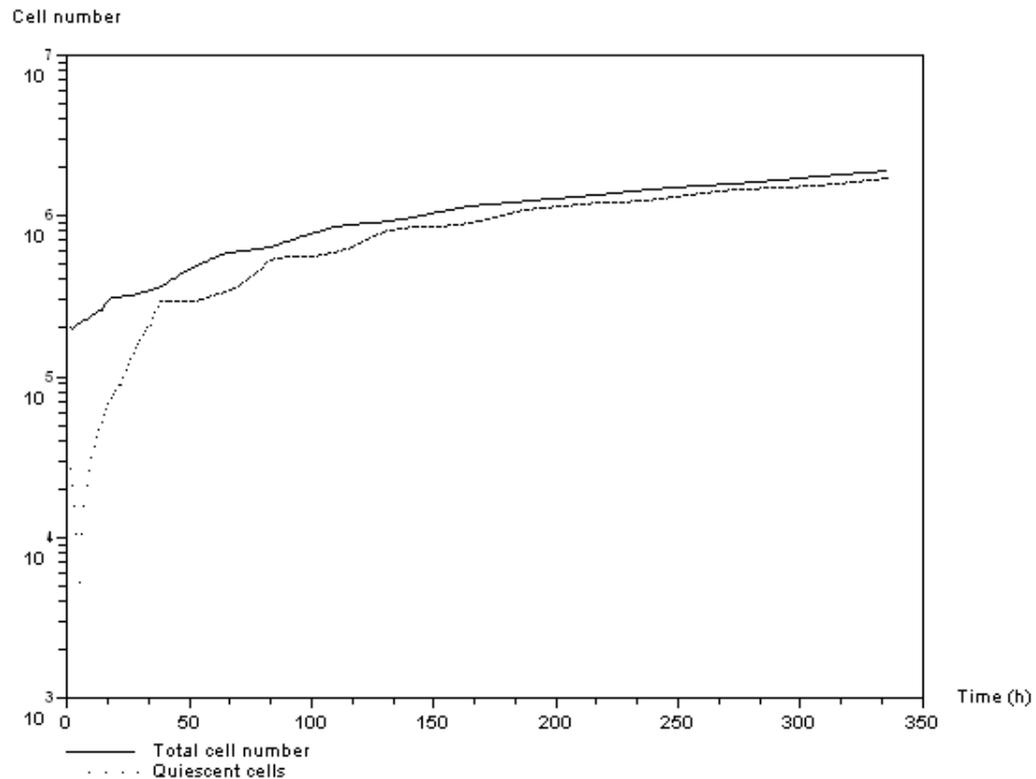
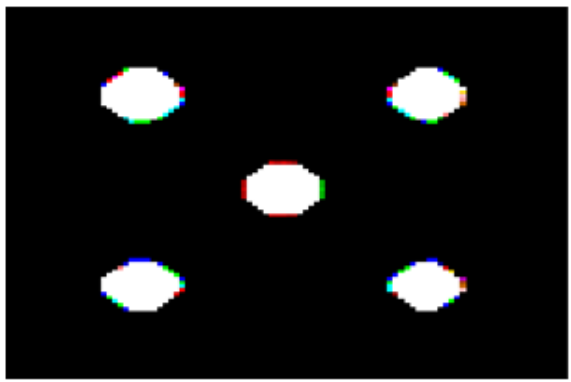
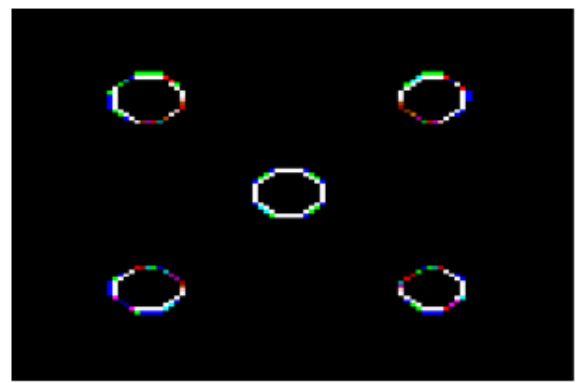

#### Figure 6
***APC*-dependent growth regulation**. Top: Evolution of the number of quiescent cells and total number of cells over time (log plot). Cell population is regulated through *APC* activation owing to overpopulation. Total cell number (continuous line) and number of quiescent cells (dotted line). Bottom: Snapshots of cells within the computational domain during simulation ($t$ = 100 h). Left: Total cell number. Right: Mitotic cells are only in the outer region of the tumor masses. Cells at the core are quiescent through *APC* activation due to overpopulation.

tion of new cells becoming quiescent (see the late phase of the curves Figure 6). The two snapshots presented at the bottom of Figure 6 show the spatial distribution of all cells (left), and that of mitotic cells only (right). Mitotic cells are situated on the outer region owing to overpopulation in the central parts of the clusters.

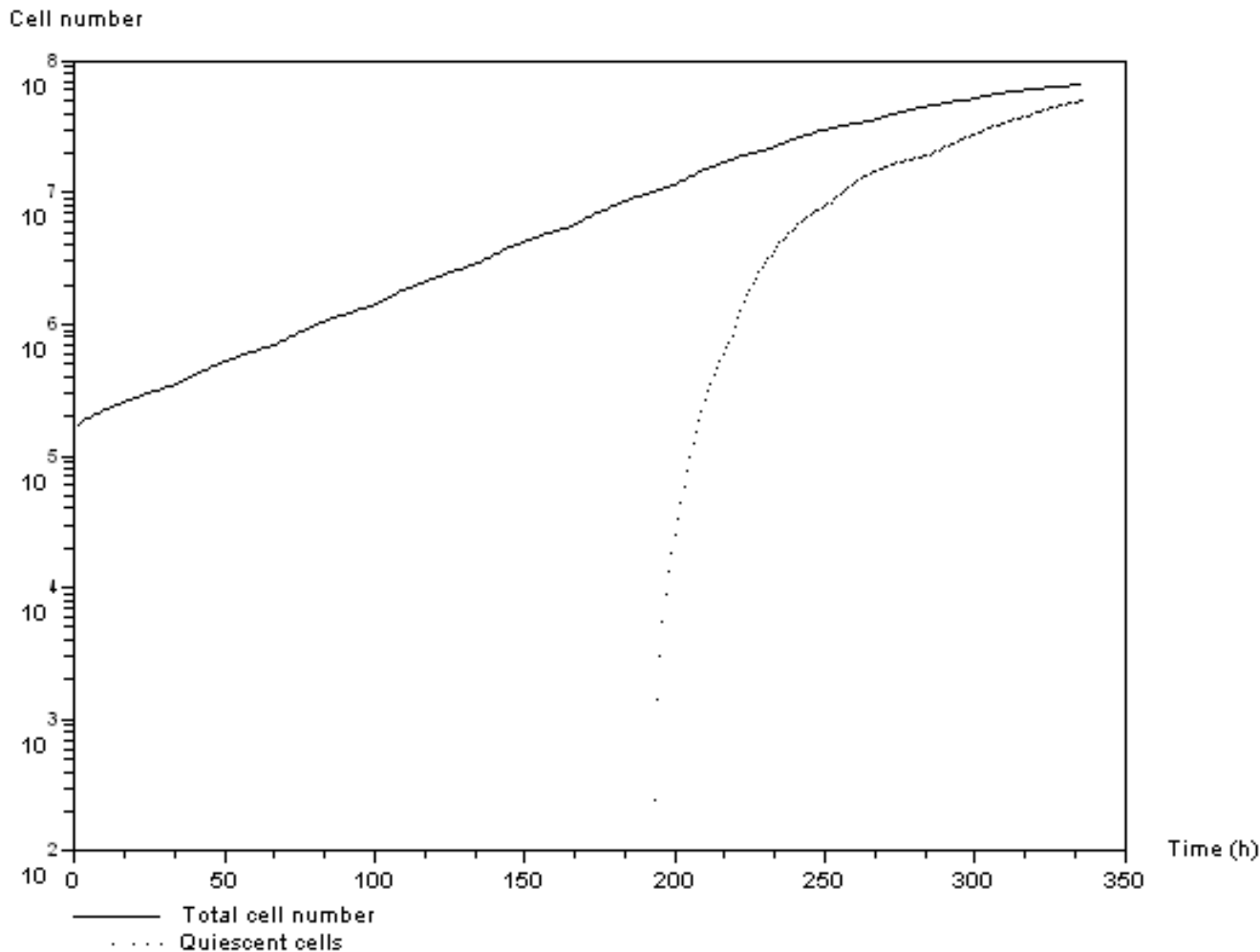

### Figure 7
***SMAD/RAS-dependent growth regulation***. Evolution of the number of quiescent cells and total number of cells over time (log plot). An *APC* mutated cell population is regulated through *SMAD/RAS* activation due to hypoxia. Total number of cells (continuous line) and number of quiescent cells (dotted line).

*SMAD/RAS-dependent growth regulation*
Figure 7 shows the time courses of total cell number and quiescent cell number. In this figure, cells are *APC* mutated and the growth regulation is controlled by *SMAD/RAS* signaling, which has been activated by hypoxia. Before hypoxia, cell population growth is exponential and becomes more linear as the anti-growth signals start.

Figure 8 shows the evolution of the number of spatial units in the computational domain co-opted by the two regulation signals. The overpopulation and hypoxia signal curves can be related to the evolution of the quiescent cells from Figure 6 and Figure 7 respectively. Figure 8 reveals the difference in evolution between the hypoxia and overpopulation signaling within the computational domain. The first oscillating growth phase depicted in Figure 6 is caused by the step-by-step evolution of the overpopulation signal activation. Hypoxia activation depicted in Figure 8 appears later and displays a sharp increase. While the overpopulation signal is local – it depends only on the local conditions – activation of the hypoxia signal is due to non-local effects. Oxygen absorbed by the cells at a particular position is not available for neighboring cells.

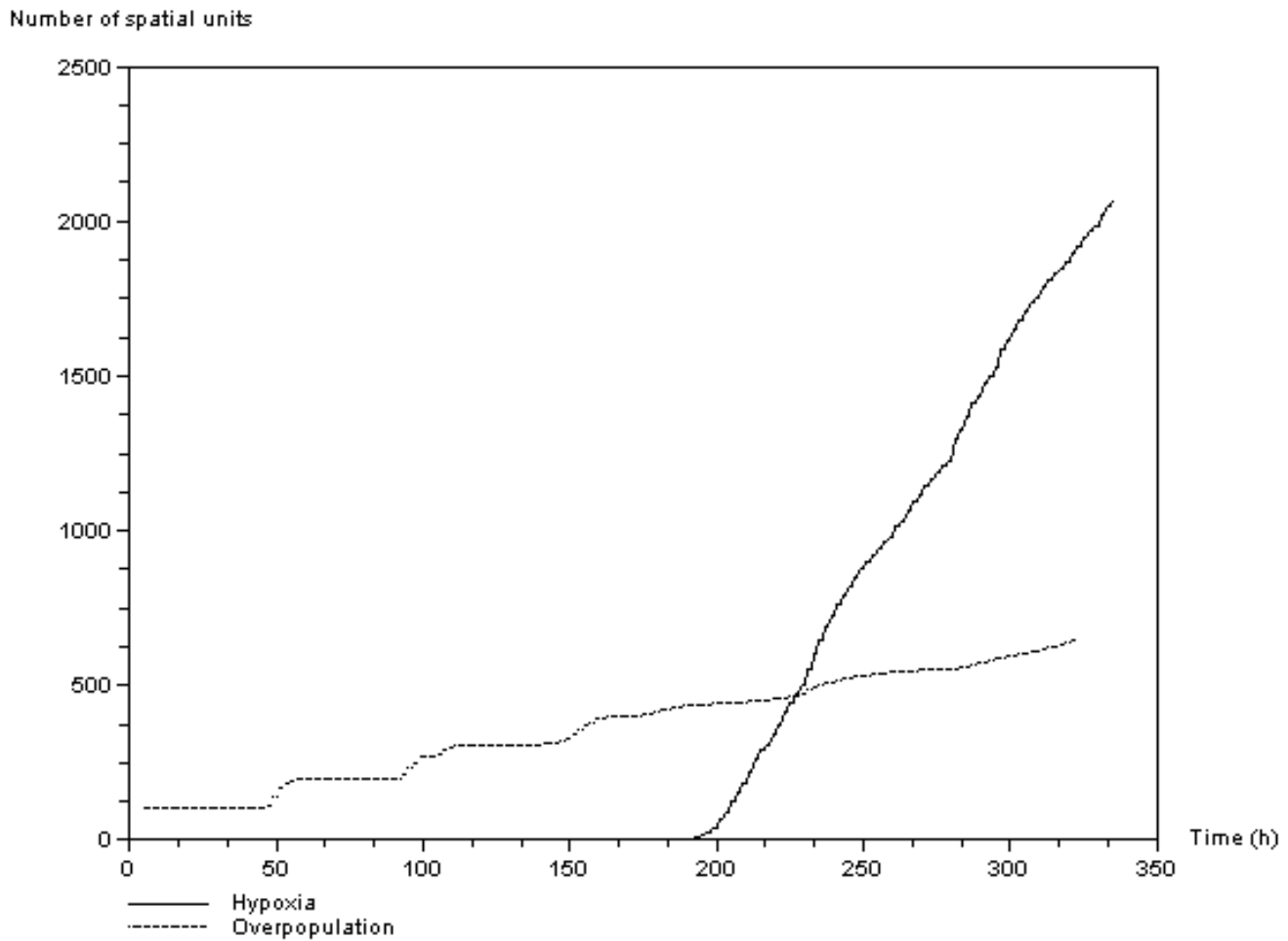

**Figure 8**
**Anti-growth signals**. Number of spatial units of the computation domain co-opted by the two regulation signals. The two curves show the activation of the hypoxia signal (continuous line) and the overpopulation signal (dashed line) over time. The vertical axis represents the number of elementary spatial units of the computational domain.

This results in regular signal propagation within the inner parts of the cell clusters as shown in the snapshots of Figure 9. Hypoxia starts from an outer area of the computational domain, i.e. areas more distant from the oxygen sources, and later occurs in the central cell cluster, where oxygen concentration is highest.

### Influence of gene-dependent growth regulation on the response to irradiation protocols

*Simulated irradiation protocols on APC and SMAD/RAS mutated tumor cells*

Figure 10 shows the evolution of the number of mutated cells going through apoptosis due to the standard irradiation protocol. In our model the treatment damages a constant fraction of mitotic cells. *APC* and *SMAD/RAS* mutated cells are not sensitive to anti-growth signals; they are in hypoxic and overpopulation conditions that lead mitotic cells to grow without regulation. Therefore the number of apoptotic cells is increased by the irradiation treatment. However, the number of apoptotic cells resulting from one treatment cycle is strictly equivalent to that induced by the previous therapeutic cycle. This is due to the difference between cell cycle duration (33 hours) and application of the treatment (24 hours).

*Simulated irradiation protocols and APC-dependent tumor growth*

When cells are sensitive to overpopulation (see growth curves Figure 6), population growth becomes linear after a first oscillating stage. Figure 11 shows the difference in efficacy between two irradiation protocols that are strictly equivalent in terms of the total dose delivered. The first is the standard protocol (dashed line), where the two doses

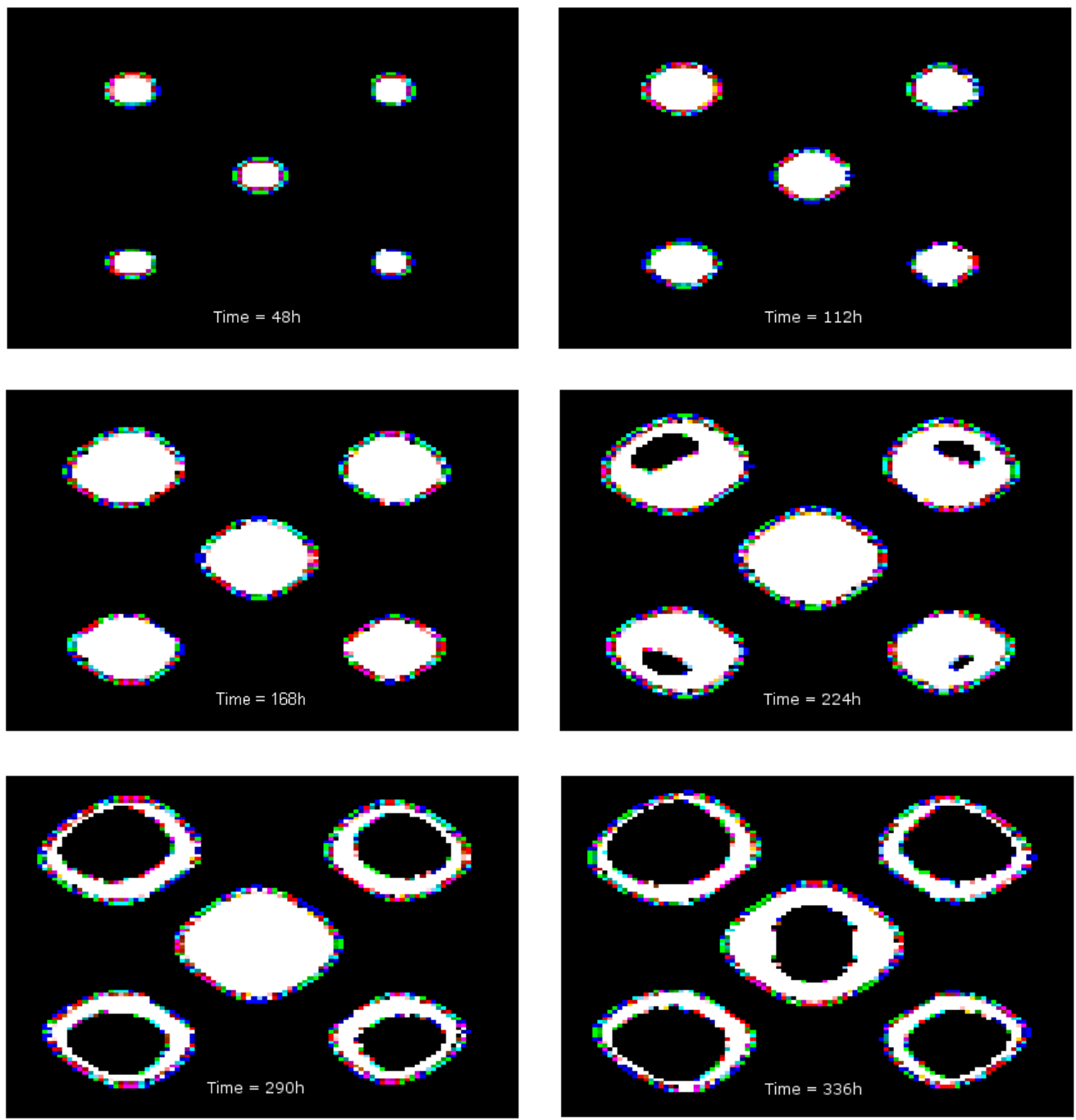

### Figure 9
**Evolution of the spatial distribution of mitotic cells**. Temporal propagation of hypoxia signal within the tumor masses. Inner black areas are cells in quiescence due to *SMAD/RAS* activation through hypoxia. The spatial distribution of mitotic cells at: top-left 48 h, top-right 112 h, middle-left 168 h, middle-right 224 h, bottom-left 290 h, bottom-right 336 h.

are delivered with a 24 h interval. The second is a heuristic approach, in which we optimized delivery of the second dose by taking account of cell cycle regulation; the second treatment is given when the number of the mitotic cells reaches a maximum. The first treatment application decreases the number of tumor cells. (Note that the dotted line in Figure 11 is hidden by the continuous line.) This also occurs in the second treatment of the heuristic proto-col. However, when the second treatment is delivered without taking growth regulation into account, i.e. standard scheduling, the efficacy is very poor (see Figure 11).

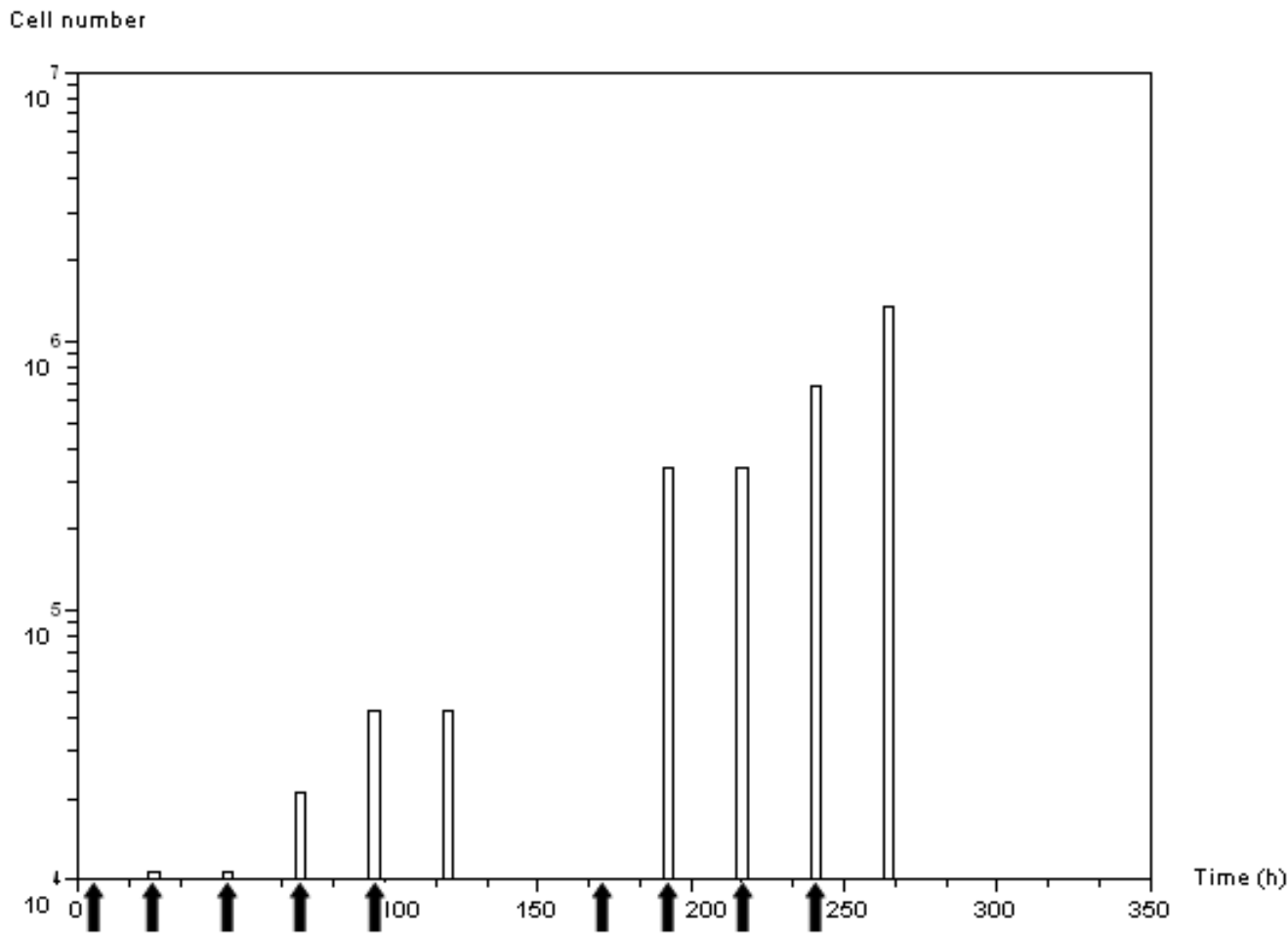

**Figure 10**
**Apoptotic activity**. Number of cells in the apoptotic phase over time when applying the standard radiotherapeutic protocol: 2 *Gy* daily. Vertical black arrows indicate treatment delivery times. Note that apoptotic activity appears at a fixed time after treatment delivery. This is the time needed for the $G_2M$ DNA-injured cells to reach the restriction point of the cell cycle (21 hours according to the model parameters).

*Simulated irradiation protocols on APC-mutated (SMAD/RAS-dependent) tumor growth regulation profiles*
Figure 12 shows the evolution of the irradiated target cell population fraction, by which we mean the time course of the mitotic fraction without irradiation, before and after activation of the hypoxia signal. As soon as the hypoxia appears, the mitotic fraction collapses. Table 1 shows the difference in simulated efficacy between two equivalent protocols in terms of total dose. The first is the standard protocol, where the 2*Gy* treatments are given daily, 5 days a week for 2 weeks, with a total dose of 20*Gy*. The second is the heuristic treatment, in which all 10 doses of 2*Gy* are given before the hypoxia signals appear. Part of the standard treatment is delivered while the tumors are becoming hypoxic (mitotic fraction falls), and this results in a decrease in efficacy. In contrast, all 10 doses in the heuristic treatment are delivered before hypoxia, which gives improved efficacy.

*Sensitivity to model parameters and initial conditions*
We study the potential influence of the choice of parameters values on the model's results. The most critical parameters to account for include:

• cell-specific radiosensitivity parameters ($\alpha_\phi$);

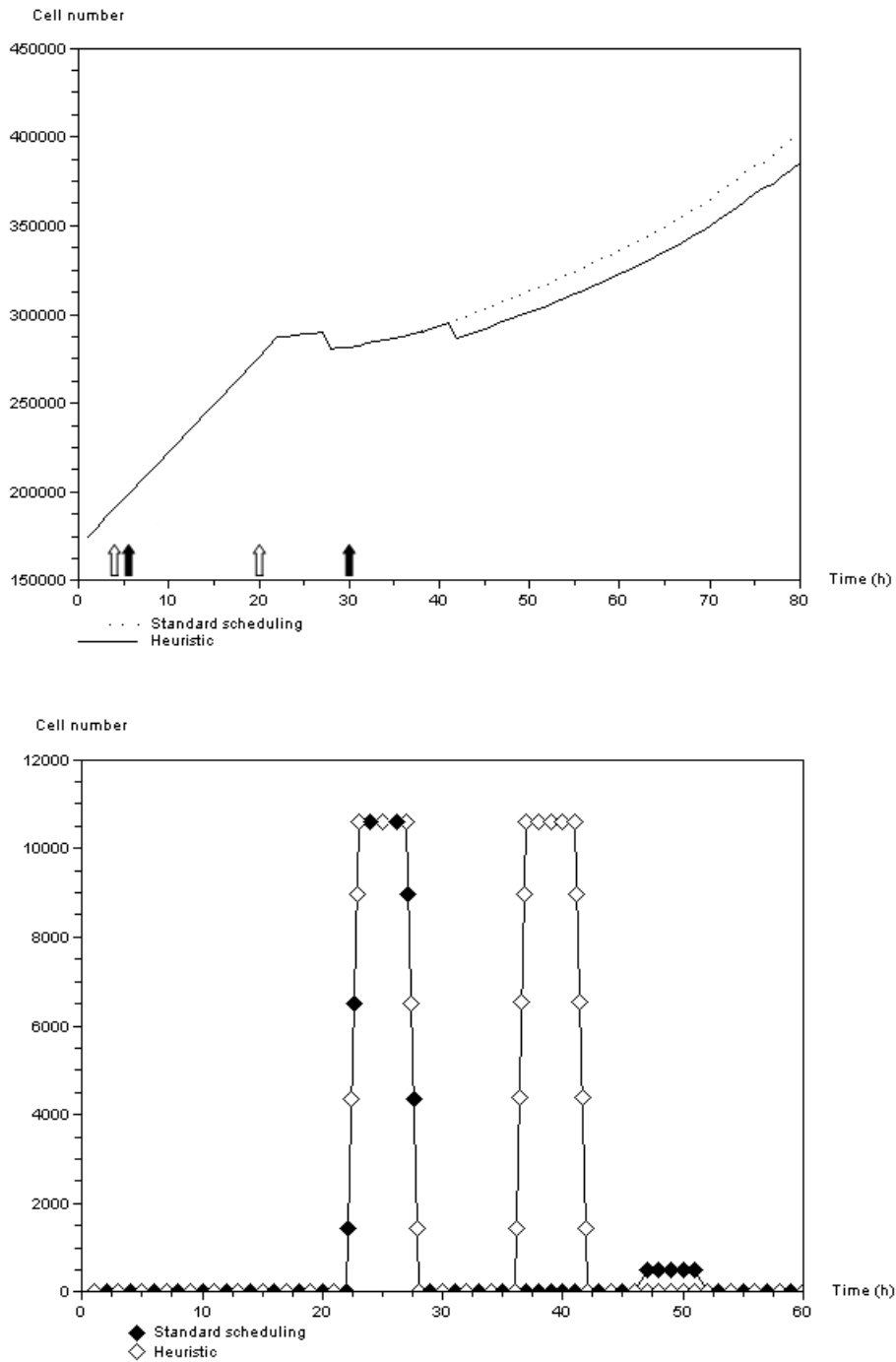

#### Figure 11
**Comparison of two radiotherapeutic protocols**. Top: Total cell number in response to standard therapeutic scheduling, i.e. 2 *Gy* applied twice within a 24 hour interval, and in response to a heuristic scheduling. Note that for the first 40 hours, the dotted line is superimposed on the continuous line since until the treatments diverge the populations are the same. Bottom: Evolution of the number of apoptotic cells due to irradiation protocols. The first treatment induces the same number of apoptotic cells. The effect of the second treatment in the standard protocol is negligible (black diamonds around time 50 h) in contrast to the heuristic approach (white diamonds pick at 40 h). Treatment delivery times are symbolized by vertical arrows: unfilled diamonds for the standard scheduling and solid diamonds for the heuristic approach.

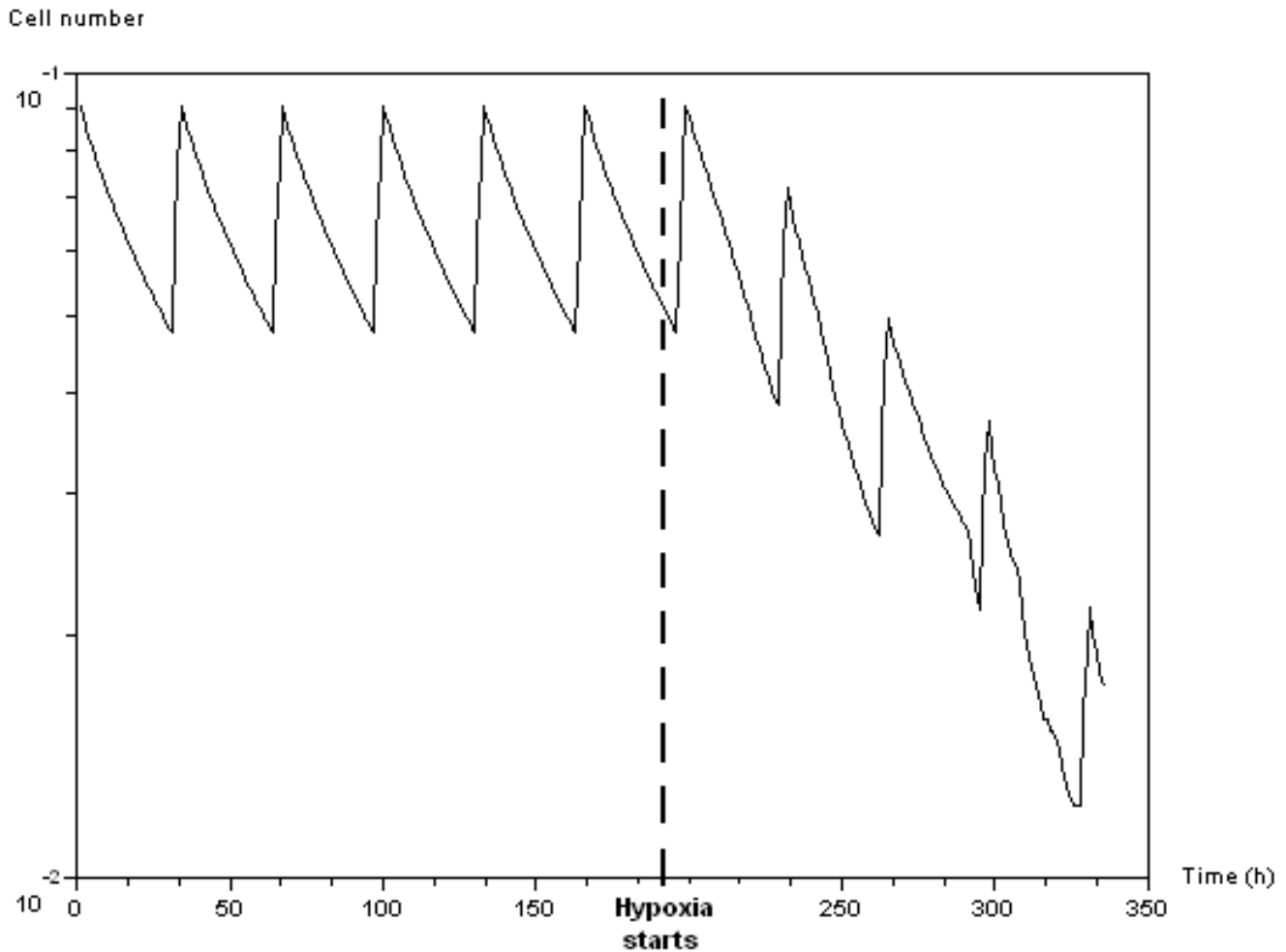

**Figure 12**
**Evolution of simulated mitotic fraction of *APC*-mutated cells over time without irradiation**. The vertical dashed line indicates the time when the hypoxia signal is activated.

• anti-growth signals, i.e. hypoxia and overpopulation, activation thresholds above which cells go into quiescence ($Th_o$ and $Th_t$);

• initial conditions, i.e. initial number of cells and spatial configurations of oxygen sources.

Treatment protocol efficacy depends directly on cell-specific radiosensitivity parameters. Figure 13 compares the evolution of total cell number over time when the standard treatment protocol is applied. Model simulations show that the standard treatment is efficient when the parameters make cells in $G_1$ phase become radiosensitive. *APC* and *SMAD/RAS* activation, which leads cells to become quiescent, is controlled by the two threshold parameters $Th_t$ and $Th_o$. Increasing $Th_t$ results in delay of the overpopulation signal, while increasing $Th_o$ speeds hypoxia activation.

Decreasing the initial number of cells has the same effect as increasing $Th_t$, while decreasing the number or the initial strength of the oxygen sources has the same effect as increasing $Th_o$. The initial configuration of tumor cells and oxygen sources is important for spatial propagation of the hypoxia signal. Indeed, Figure 9 shows a particular hypoxia propagation in the tumor cell masses that is correlated with the locations of the oxygen sources. Since $Th_t$ and $Th_o$ are merely constants, it seems that we may change

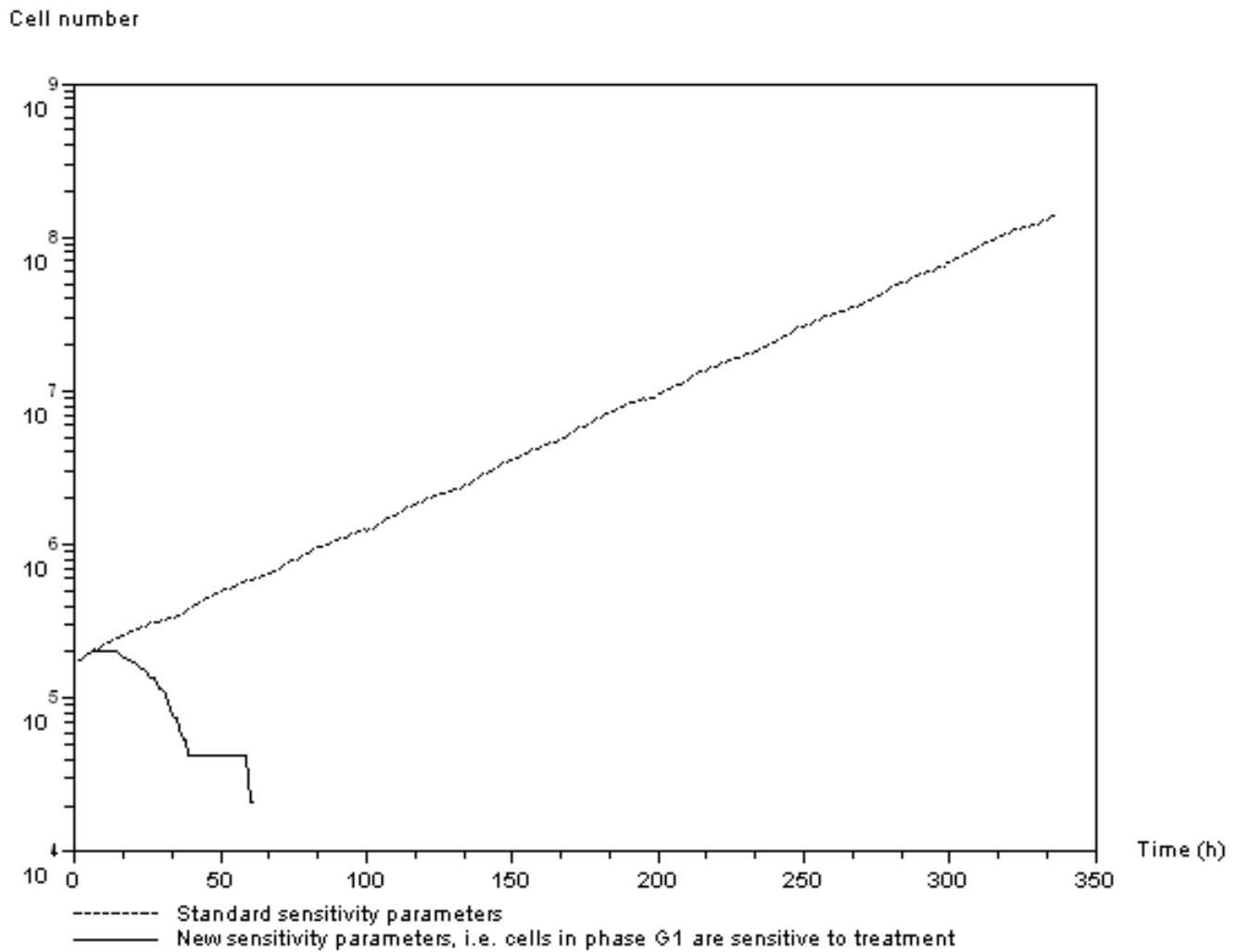

#### Figure 13
**Effect of radiosensitivity parameters on treatment efficacy**. Evolution of total cell number over time with the standard radiosensitivity parameters (continuous line), and with the suggested parameters. This shows that, with the new treatment, cells in $G_1$ phase are sensitive to the 2 *Gy* treatment dose.

the spatial configuration and size of the initial cell population and vary the oxygen sources and yet produce the same qualitative results.

Finally, Figure 14 shows the difference in evolution of the overpopulation signal over time if the initial distribution of cells in the clusters is uniform or random. The step by step evolution of overpopulation activation is softened but still exists when the cells are randomly distributed within the initial tumor masses.

## Conclusion
We have presented a multiscale model of cancer growth and examined the qualitative response to radiotherapy. The mathematical framework includes a Boolean description of a genetic network relevant to colorectal oncogenesis, a discrete model of the cell cycle and a continuous macroscopic model of tumor growth and invasion. The basis of the model is that the sensitivity to irradiation depends on cell cycle phase and that DNA damage is proportional to the radiation dose. Anti-growth regulation signals such as hypoxia and overpopulation activate the *SMAD/RAS* and *APC* genes, respectively, and inhibit proliferation through cell cycle regulation.

Simulation results show the different features of the anti-growth signal activation and propagation within the tumor (see Figure 8). The overpopulation signal mediated by the *APC* gene initially induces oscillatory growth owing to a combination of proliferating and quiescent cells (see

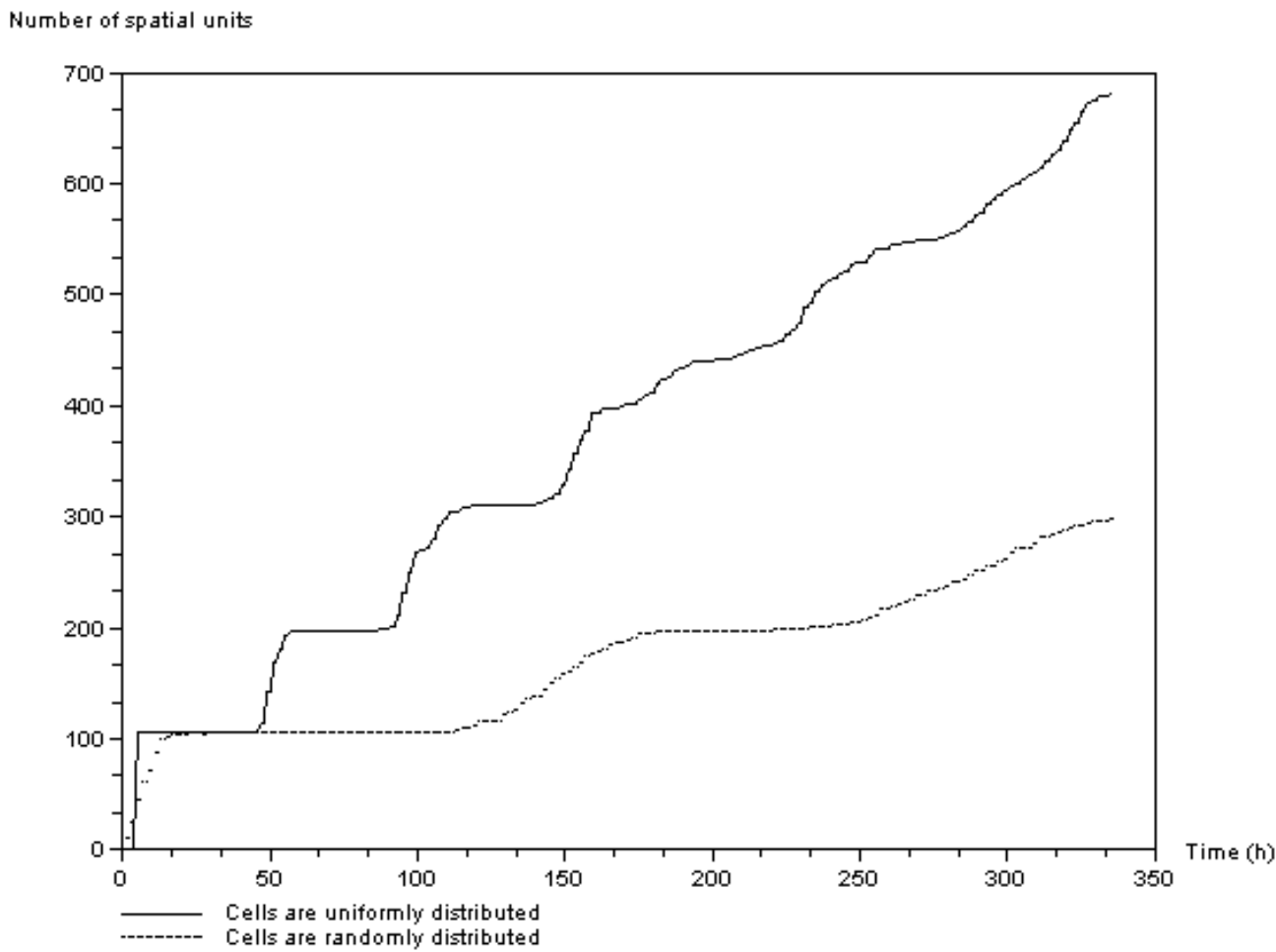

**Figure 14**
**Effect of cell distribution within the initial cell clusters on overpopulation**. The vertical axis is the number of elementary spatial units of the computational domain. Here we show the difference between evolution of the overpopulation signal over time when cells in the clusters are initially distributed uniformly or randomly. The evolution of overpopulation activation is softened but still exists when cells are randomly distributed within the initial tumor masses.

Figure 6). Because of its non-local effect, the hypoxia signal mediated by genes *SMAD/RAS* appears later but develops quickly within the tumor masses, and leads the mitotic fraction to collapse (see Figures 11 and 14). These features make the evolution of the number of quiescent cells and thus the efficacy of irradiation protocols depend on the type of anti-growth signals to which the tumors are exposed. Figure 11 and Table 1 show that efficacy could be improved, without increasing radiation doses, by planning schedules that take account of the features of tumor growth through cell cycle regulation.

The proposed framework emphasizes the significant role of gene-dependent cell-cycle regulation in the response of tumors to radiotherapy. Clinical studies have recognized *p53* status as a major predictive factor for the response of rectal cancer to irradiation. Nevertheless, some results encourage investigation of other different factors [46]. In particular, it has been suggested that macroscopic factors such as hypoxia and tumor volumes are important [47]. The present modeling framework integrates these factors through cell cycle regulation and allows consideration of other factors at the genetic, cellular or tissue level.

Some modeling assumptions must be discussed. We chose a continuous approach that provides cell density rather than actual cell number. This assumes that the region of interest is large since we have restricted our analysis to late-stage tumor development. We have not considered cell shape, which has been shown to be important for the correct description of growth control processes [48]. Individual-based models of cell movement, e.g. the Potts model [49,50] and the Langevin model [51], would improve our approach. We reduced the system to two dimensions. A three-dimensional tumor model could reveal new factors in the dynamics.

The aim of this study is to understand the qualitative effect of therapeutic protocols on colorectal cancer. Our analysis raises some interesting points about the influence of anti-growth regulation signals and genetic pathways on the efficacy of the standard protocol. Efforts have been made to improve the LQ model by taking into account multiple factors such as tumor volume and repopulation between treatment cycles [52]. However, we have produced a multiscale model that is more realistic and demonstrated its use in comparing efficacy of treatment protocols.

## Authors' contributions
BR designed the mathematical multiscale model and simulated it to investigate the role of cell cycle regulation in response to irradiation treatment protocols. TC designed the macroscopic level. He implemented the advection-diffusion equations and contributed to linking the sub-models together. SS elaborated the genetic Boolean network model of colorectal oncogenesis and its implementation. He also supervised manuscript preparation and revision.

## Acknowledgements
BR is funded by the ETOILE project: "Espace de Traitement Oncologique par Ions Légers dans le cadre Européen". Part of this work was carried out during the "Biocomplexity Workshop 7" held at Indiana University (Bloomington Campus) in May 9–12, 2005. The workshop was sponsored by the National Science Foundation (Grant MCB0513693) and the National Institute of General Medical Science/National Institutes of Health (Grant R13GM075730). BR is very grateful for the hospitality of the Indiana University School of Informatics and the Biocomplexity Institute during his visit May 8–14. The authors wish to acknowledge particularly the two referees for their useful comments; Professor Jean-Pierre Boissel and François Gueyffier for manuscript review; Professor Emmanuel Grenier, Dr Didier Bresch, and Nicolas Voirin for their valuable advice regarding model implementation; and Dr Ramon Grima and Edward Flach for critical comments.